\documentclass[12pt]{article}

\usepackage{amssymb}
\usepackage{amsmath}
\usepackage{amscd}
\usepackage{latexsym}
\usepackage{graphicx}

\usepackage{cite}

\newcommand{\be}{\begin{equation}}
\newcommand{\ee}{\end{equation}}

\newcommand{\br}{{\bf r}}

\newcommand{\vp}{\varphi}
\newcommand{\ep}{\varepsilon}
\newcommand{\al}{\alpha}

\newcommand{\gm}{\gamma}
\newcommand{\om}{\omega}

\newcommand{\Gm}{\Gamma}
\newcommand{\ra}{\rightarrow}

\begin{document}

\begin{center}

{\Large{\bf Self-similar extrapolation of nonlinear problems from small-variable 
to large-variable limit} \\ [5mm]

V.I. Yukalov$^{1,2,*}$, E.P. Yukalova$^3$} \\ [3mm]

$^1${\it Bogolubov Laboratory of Theoretical Physics, \\
Joint Institute for Nuclear Research, Dubna 141980, Russia \\ [2mm]

$^2$Instituto de Fisica de S\~ao Carlos, Universidade de S\~ao Paulo, \\
CP 369,  S\~ao Carlos 13560-970, S\~ao Paulo, Brazil   \\ [2mm]

$^3$Laboratory of Information Technologies, \\
Joint Institute for Nuclear Research, Dubna 141980, Russia } \\ [3mm]

$^*${\bf Corresponding author e-mail}: yukalov@theor.jinr.ru

\end{center}

\vskip 2cm

\begin{abstract}   

Complicated physical problems usually are solved by resorting to perturbation theory 
leading to solutions in the form of asymptotic series in powers of small parameters.  
However, finite, and even large values of the parameters often are of the main physical 
interest. A method is described for predicting the large-variable behaviour of solutions
to nonlinear problems from the knowledge of only their small-variable expansions. The 
method is based on self-similar approximation theory resulting in self-similar factor 
approximants. The latter can well approximate a large class of functions, rational, 
irrational, and transcendental. The method is illustrated by several examples from 
statistical and condensed matter physics, where the self-similar predictions can be 
compared with the available large-variable behaviour. It is shown that the method allows
for finding the behavior of solutions at large variables when knowing just a few terms
of small-variable expansions. Numerical convergence of approximants is demonstrated.     
\end{abstract}

\vskip 1cm

{\parindent=0pt

{\it Keywords}: Nonlinear physical problems; asymptotic series; self-similar 
extrapolation; large-variable behaviour; numerical convergence

\vskip 1cm

PACS numbers: 02.30.Mv, 02.60.-x }

\newpage

\section{Introduction}

There exist numerous problems in statistical physics, condensed-matter theory, and 
atomic physics, where the sought characteristic is defined by so complicated equations 
that it can be found only in the region of small parameters, being given by an 
asymptotic expansion. However, one may need this characteristic in the region of 
finite parameters and, moreover, one often needs the knowledge of the characteristic 
behaviour at very large parameters tending to infinity. In the present paper, we 
concentrate on the latter, the most difficult, problem of defining the large-variable 
behaviour of functions from the knowledge of only their small-variable asymptotic 
expansions. 

It is clear that Pad{\'e} approximants cannot describe the large-variable behaviour 
of the sought quantity for which the large-variable limit is not known \cite{Baker_1}. 
This is because a Pad{\'e} approximant $P_{M/N}(x)$ describes the large-variable 
behaviour as proportional to $x^{M-N}$, with an integer power, while, generally, 
this power can be irrational. Moreover, for the same problem one can define the 
whole table of Pad{\'e} approximants, with different exponents $M - N$ in the term 
$x^{M-N}$, so that no concrete prediction can be made.  

In addition, there exists the well known problem of incompatibility of small-variable 
and large-variable expansions. This is when the small-variable expansion is in powers 
of $x^\alpha$, while the large-variable expansion is in powers of $x^\beta$, such 
that $\alpha/\beta$ is an irrational number, because of which there is no such a 
change of the variable that could allow for sewing the small-variable and large-variable 
limits. Several other problems in using  Pad{\'e} approximants have been discussed 
in \cite{Baker_1,Baker_2,Gluzman_3,Gluzman_4}.

The other way that could make it possible to guess the large-variable behaviour is 
the method of Borel transforms \cite{Hardy_5,Weinberg_6}. But this method requires 
the knowledge of high-order terms of a small-variable expansion, which is rarely 
available. And it is not applicable to the cases where just a few terms of the 
small-variable expansion are provided. Several other methods of constructing 
approximate solutions to nonlinear problems are reviewed in Ref. \cite{He_7}. 

In the present paper, we describe an original method that makes it straightforward 
to get reasonably accurate estimates for the large-variable limits of the sought 
solutions from the knowledge of only their small-variable expansions. The method is 
based on self-similar approximation theory that was advanced in Refs. \cite{Yukalov_8}  
and described in all details in the previous publications 
\cite{Yukalov_9,Yukalov_10,Yukalov_11,Yukalov_12,Yukalov_13}. The efficiency of the
self-similar approximation theory stems from its use of the ideas of dynamic theory and 
renormalization-group theory \cite{Halmos_14,Ladyzhenskaya_15,Lichtenberg_16,Yukalov_17}, 
as well as of optimal control theory \cite{Foulds_18,Hocking_19}. The use of 
renormalization-group theory is known to provide physically reasonable solutions to 
problems with divergencies \cite{Adhikari_54,Adhikari_55}. 

In the frame of the self-similar approximation theory, the small-variable asymptotic 
expansions can be transformed into self-similar factor approximants, extrapolating the 
given asymptotic expansions to the finite values of their variables 
\cite{Yukalov_20,Gluzman_21,Yukalov_22}. Here we demonstrate that the method allows us to 
extrapolate small-variable series, not merely to finite variables, but it makes it 
possible to predict the large-variable limits of the sought solutions.  
 
Since the justification for the self-similar approximation theory has been expounded
in our previous papers, we do not repeat here the published earlier material. A brief 
account of the basic points of the self-similar approximation theory can be found in the 
recent article \cite{Gluzman_23}. Here we only briefly recall the main steps in using the 
method of self-similar factor approximants \cite{Yukalov_20,Gluzman_21,Yukalov_22}.

By treating  the cases, where many terms of the asymptotic expansions are known, 
we prove that the method is numerically convergent. Moreover, we demonstrate that it 
predicts rather good estimates even for the cases, where just a few terms are available. 
We show this by several examples from statistical physics, condensed-matter theory, and 
atomic physics, in which the large-variable limits are known, thus, providing us 
the possibility of estimating the accuracy of our predictions.

\section{Self-similar factor approximants}

Let the bare approximants, valid at asymptotically small $x \ra 0$, have the form
\be
\label{25}
f_k(x) = f_0(x) \left( 1 + \sum_{n=1}^k a_n x^n \right ) \;   ,
\ee
where $f_0(x)$ is a given function. The sum in the brackets is usually divergent 
for finite values of $x$ and makes no sense for such $x$. Moreover, in many cases 
it is important to know the behaviour of the sought function $f(x)$ at asymptotically 
large $x \ra \infty$. This is exactly the problem we shall concentrate on.

First, by the fundamental theorem of algebra \cite{Lang_24}, it is known that 
a polynomial of any degree of one real variable over the field of real numbers can 
be split in a unique way into a product of irreducible first-degree polynomials over 
the field of complex numbers. This means that the finite series (\ref{25}) can be 
represented as the product
\be
\label{26}
 f_k(x) = f_0(x) \prod_j \left( 1 +  b_j x \right ) \; ,
\ee
with $b_j$ expressed through $a_n$.  

Self-similar properties can be explicitly incorporated into a map employing fractal 
transforms \cite{Barnsley_25,Yukalov_26}. Keeping this in mind, we introduce control 
functions $s_k$ and $u_k$ by the fractal transform  
\be
\label{27}
 \hat T[s,u] f_k(x) = x^{s_k} f_k(x) + u_k \; .
\ee
Then following the scheme, described in Refs. \cite{Yukalov_20,Gluzman_21,Yukalov_22}, 
we obtain the self-similar factor approximants
\be
\label{28}
f_k^*(x) = f_0(x) \prod_{j=1}^{N_k} \left( 1 +  A_j x \right )^{n_j} \;  ,
\ee
with $A_j$ and $n_j$ playing the role of control parameters. The number of factors 
$N_k$ equals $k/2$ for even $k$ and $(k + 1)/ 2$ for odd $k$. If the factor approximant 
(\ref{28}) represents the sought function, then their asymptotic expansions should 
coincide. Therefore, the parameters $A_j$ and $n_j$ have to be chosen so that the 
asymptotic expansion of (\ref{28}), of order $k$, be equal to the asymptotic form 
(\ref{25}), that is,
\be
\label{29}
 f_k^*(x) \simeq f_k(x) \qquad ( x \ra 0 ) \; .
\ee
This condition yields the equations
\be
\label{30}
 \sum_{j=1}^{N_k} n_j A_j^n = D_n \qquad ( n = 1,2,\ldots, k) \;  ,
\ee
where
$$
D_n \equiv \frac{(-1)^{n-1}}{(n-1)!} \; \lim_{x\ra 0} \; 
\frac{d^n}{dx^n} \; \ln\left( 1 + \sum_{m=1}^n a_m x^m \right ) \; .
$$
 
When $k$ is even, and $N_k = k/2$, we have $k$ equations for $k$ unknown parameters 
$A_j$ and $n_j$, uniquely defining these parameters \cite{Yukalov_27}. But if $k$ 
is odd, and $N_k = (k + 1)/2$, we have $k$ equations for $k + 1$ parameters. Then, 
to make the system of equations complete, it is required to add one more condition. 
One possibility, based on scaling arguments \cite{Yukalov_27}, could be to set one 
of $A_j$ to one, say fixing $A_1 = 1$. Although this condition is acceptable and 
gives the results close to the nearest even-order approximants, we shall not use 
it below, dealing mainly with uniquely defined even orders.   

It is important to emphasize that there exists a large class of functions that are 
exactly reproducible by the factor approximants \cite{Yukalov_20,Gluzman_21,Yukalov_22}. 
This class consists of the functions that can be reduced to the form
\be
\label{31}
  R_M(x) = \prod_{j=1}^M P_{m_j}^{\al_j}(x) \; ,
\ee
where $P_m$ are polynomials of degree $m$ and $\alpha_j$ are any real numbers or 
complex-valued numbers entering by pairs with their complex-conjugate, which makes
the function $R_M(x)$ real. A function from this class is exactly reproducible 
by factor approximants of order 
$$
k \geq \sum_{j=1}^M m_j \; .
$$
Thus the class of exactly reproducible functions includes rational, as well as 
irrational functions. Moreover, it also includes transcendental functions that are 
the limits of form (\ref{31}). For instance, the exponential function is exactly 
reproducible for any order $k \geq 2$, due to the limit 
$$
 \lim_{b\ra 0} P_1^{1/b}(x) = a e^{x/a} \; \qquad P_1(x) = a + bx \;  .
$$
 
As is stressed above, in the present paper, we are interested in the possibility 
of predicting the large-variable behaviour. Let the given function $f_0$ behave as
\be
\label{32}
 f_0(x) \simeq A x^\al \qquad ( x \ra \infty) \;  .
\ee
Then the self-similar factor approximant (\ref{28}) for large $x$ is
\be
\label{33}
 f_x^*(x) \simeq B_k x^{\gm_k} \qquad ( x \ra \infty) \;  ,
\ee
with the amplitude
\be
\label{34}
B_k = A \prod_{j=1}^{N_k} A_j^{n_j}
\ee
and the exponent
\be
\label{35}
\gm_k = \al + \sum_{j=1}^{N_k} n_j \; .
\ee
In those cases, where the large-variable asymptotic behaviour of the sought 
function is known, say being
\be
\label{36}
f(x) \simeq B x^\gm \qquad ( x \ra \infty) \;  ,
\ee
we can determine the accuracy of our prediction (\ref{33}) by calculating the 
percentage errors
\be
\label{37}
\ep(B_k) \equiv \frac{B_k-B}{B} \times 100\% \; , \qquad 
\ep(\gm_k) \equiv \frac{\gm_k-\gm}{\gm} \times 100 \% \;   .
\ee
And if the exact large-variable asymptotic behaviour is not available, then the 
accuracy can be estimated by the difference between the subsequent values of the
sought quantity.

\section{Convergence of factor approximants}

In order to show that self-similar factor approximants provide convergent series, 
we need to consider the problems, where a number of terms are available. Below we 
study several such cases.

\subsection{Partition function of zero-dimensional oscillator}

The standard problem that one almost always considers as a probe of numerical 
procedures is the partition function of zero-dimensional oscillator
\be
\label{38}
Z(g) = \frac{1}{\sqrt{\pi}} \; \int_{-\infty}^\infty e^{-\vp^2-g\vp^4}\; d\vp \; .
\ee
Expanding this function in powers of the coupling parameter yields the series
\be
\label{39}
 Z_k(g) =\sum_{n=0}^k a_n g^n \qquad ( g\ra 0) \;  ,
\ee
with the coefficients
$$
a_n = \frac{(-1)^n}{\sqrt{\pi}\; n!} \; \Gm\left( 2n + \frac{1}{2}\right ) \; .
$$
The obtained series have the typical structure of many problems in condensed-matter 
theory and statistical physics, with factorially growing coefficients signaling 
divergence of the series for any finite value of $g$. The strong-coupling limit is known 
to be
\be
\label{40}
Z(g) \simeq 1.022765 g^{-1/4} \qquad ( g \ra \infty) \; .
\ee
Defining factor approximants (\ref{28}) for series (\ref{39}), we consider the limit
\be
\label{41}
 Z_k^*(g) \simeq B_k g^{\gm_k} \qquad ( g \ra \infty) \;  .
\ee
The results for the amplitude $B_k$ and exponent $\gamma_k$, together with their 
percentage errors (\ref{37}) are listed in Table 1. We see the monotonic decrease of 
the errors, which means the evident numerical convergence to the exact values for both 
the amplitude and exponent.

\begin{table}
\caption{Amplitudes and exponents, with their percentage errors, for the 
strong-coupling limit of the zero-dimensional partition function, predicted 
by self-similar factor approximants.}
\vskip 3mm
\label{Table 1}
\renewcommand{\arraystretch}{1.25}
\centering
\begin{tabular}{|c|c|c|c|c|}
\hline
$k$ & $B_k$  &  $\ep(B_k)\%$  &  $\gm_k$  &  $\ep(\gm_k)\%$ \\ \hline
2   & 0.823  &  $-$19.5       &  $-$0.090 &  $-$62.5    \\
4   & 0.806  &  $-$21.2       &  $-$0.129 &  $-$48.4     \\
6   & 0.806  &  $-$21.2       &  $-$0.148 &  $-$40.6     \\
8   & 0.810  &  $-$20.8       &  $-$0.161 &  $-$35.6     \\
10  & 0.814  &  $-$20.4       &  $-$0.170 &  $-$32.0    \\
12  & 0.819  &  $-$19.9       &  $-$0.178 &  $-$29.3    \\   
14  & 0.824  &  $-$19.4       &  $-$0.182 &  $-$27.1    \\ 
16  & 0.828  &  $-$19.0       &  $-$0.187 &  $-$25.4    \\ \hline
\end{tabular}
\end{table}

\subsection{Ground-state energy of anharmonic oscillator}

The other typical touchstone for analyzing numerical methods is the anharmonic 
oscillator with the Hamiltonian
\be
\label{42}
H = - \; \frac{1}{2} \; \frac{d^2}{dx^2} + \frac{1}{2} \; x^2 + g x^4 \; ,
\ee
in which $x \in (-\infty, \infty)$ and $g \in [0, \infty)$. The ground-state energy 
of this Hamiltonian can be found \cite{Bender_28,Hioe_29} as an asymptotic expansion 
in powers of the coupling parameter $g$,
\be
\label{43}
 E(g) \simeq \frac{1}{2} + \sum_{n=1}^k a_n g^n \qquad ( g \ra 0 ) \; .
\ee
The coefficients $a_n$ can be found in \cite{Bender_28,Hioe_29}, with the first 
of them being
$$
a_1 = 0.75, \qquad a_2 = -2.625, \qquad a_3 = 20.8125, \qquad a_4 = -241.2890625,
$$
$$
a_5 = 3580.98046875, \qquad a_6 = -63982.8134766, \qquad a_7 = 1329733.72705, 
$$
$$
a_8 = -31448214.6928, \qquad a_9 = 833541603.263, \qquad 
a_{10} = -24478940702.8 \; . 
$$
The strong-coupling limit is
\be
\label{44}
 E(g) \simeq 0.667986 g^{1/3} \qquad ( g \ra \infty) \;  .
\ee

Constructing factor approximants and considering their strong-coupling limit
\be
\label{45}
E_k^*(g) \simeq B_k g^{\gm_k} \qquad ( g \ra \infty ) \; ,
\ee
we find the amplitudes and exponents, with their percentage errors, listed in 
Table 2. Again we observe the monotonic decrease of the errors, implying the 
evident numerical convergence. 
 
\begin{table}
\caption{Amplitudes and exponents, with the related percentage errors, for the 
strong-coupling limit of the one-dimensional anharmonic oscillator ground-state 
energy predicted by self-similar factor approximants.}
\vskip 3mm
\label{Table 2}
\renewcommand{\arraystretch}{1.25}
\centering
\begin{tabular}{|c|c|c|c|c|}
\hline
$k$ & $B_k$  &  $\ep(B_k)\%$  &  $\gm_k$  &  $\ep(\gm_k)\%$ \\ \hline
2   & 0.729  &  9.2       &  0.176 &  $-$47.0    \\
4   & 0.755  &  13.1      &  0.231 &  $-$30.6     \\
6   & 0.756  &  13.2      &  0.257 &  $-$22.9     \\
8   & 0.752  &  12.6      &  0.272 &  $-$18.4     \\
10  & 0.748  &  11.9      &  0.282 &  $-$15.5    \\
12  & 0.743  &  11.2      &  0.289 &  $-$13.4    \\   
14  & 0.739  &  10.7      &  0.294 &  $-$11.9    \\ 
16  & 0.736  &  10.2      &  0.298 &  $-$10.7    \\ \hline
\end{tabular}
\end{table}

\subsection{Mittag-Leffler function in kinetic equations} 

Mittag-Leffler functions appear naturally in the solutions of fractional order 
integral and differential equations describing fractional generalizations of kinetic 
equations, random walks, L\'{e}vy flights, and super-diffusive transport. Here we 
consider the Mittag-Leffler function
\be
\label{46}
E(x) = e^{x^2} {\rm erfc}(x) = 
e^{x^2} \; \frac{2}{\sqrt{\pi}} \int_x^\infty e^{-t^2} \; dt \; .
\ee
It is straightforward to find the function expansion of arbitrary order in powers 
of $x$, with the low orders being
\be
\label{47}
E(x) \simeq 1 \; - \; \frac{2}{\sqrt{\pi}} \; x \; + \; x^2 \; - \;
\frac{4}{3\sqrt{\pi}} \; x^3 \; + \; \frac{x^4}{2} \qquad ( x \ra 0 ) \; .
\ee
And the large-variable limit is 
\be
\label{48}
 E(x) \simeq \frac{1}{\sqrt{\pi}} \; x^{-1} \qquad ( x \ra \infty) \; .
\ee
This function is not easy to approximate, since it consists of two different 
functions each producing its own expansion. To catch at least a couple of the 
expansion terms of each of these two functions, one needs to consider at least the 
fourth-order expansion of the whole product. 

Constructing factor approximants, we find the amplitudes and exponents at large $x$,
\be
\label{49}
E_k^*(x) \simeq B_k x^{\gm_k} \qquad ( x \ra \infty) \;  ,
\ee
predicted by these expressions, which are shown in Table 3, together with their 
percentage errors. At the beginning, there are oscillations in the values of the errors 
for the amplitude, but from sixth order, the errors monotonically decrease. The 
convergence is quite fast, reaching in the twelfth order the error of only $3 \%$ 
for the amplitude $B_k$ and $0.7 \%$ for the exponent.   

\begin{table}
\caption{Amplitudes and exponents, with their percentage errors, for the 
large-variable limit of the Mittag-Leffler function, predicted by self-similar 
factor approximants.}
\vskip 3mm
\label{Table 3}
\renewcommand{\arraystretch}{1.25}
\centering
\begin{tabular}{|c|c|c|c|c|}
\hline
$k$ & $B_k$  &  $\ep(B_k)\%$  &  $\gm_k$  &  $\ep(\gm_k)\%$ \\ \hline
4   & 0.209  &  $-$62.9.      &  $-$0.618 &  $-$38.2    \\
6   & 0.926  &  64.1          &  $-$1.160 &  16.0     \\
8   & 0.457  &  $-$18.9       &  $-$0.939 &  $-$6.1     \\
10  & 0.612  &  8.4           &  $-$1.02  &  2.2     \\
12  & 0.548  &  $-$2.9        &  $-$0.993 &  $-$0.7    \\   \hline
\end{tabular}
\end{table}

\subsection{Massive Schwinger model in Hamiltonian lattice theory}

Massive Schwinger model \cite{Schwinger_30} is the lattice model imitating quantum 
theory in $1+1$ space-time dimensions. This model exhibits several phenomena that are 
typical of quantum chromodynamics, such as confinement, chiral symmetry breaking with 
an axial anomaly, and a topological vacuum, which makes it, probably, the simplest 
nontrivial gauge theory \cite{Banks_31,Carrol_32,Vary_33,Adam_34,Striganesh_35}. This 
is why the Schwinger model is considered as a standard testbed for the trial of new 
techniques. Calculating the spectrum of excited states for a finite lattice, one meets 
\cite{Hamer_36} the functions of the type
\be
\label{50}
f(z) \simeq 1 + \sum_{k=1}^k a_n z^n \qquad ( z \ra 0 )
\ee
expanded in powers of the variable $z \equiv 1/(ga)^4$, where $g$ is the coupling 
parameter and $a$ is the lattice spacing. For the vector boson, one has the fastly 
growing coefficients
$$
a_1 = 2 \; , \qquad a_2 = -10 \; , \qquad a_3 = 78.66667 \; , \qquad
a_4 = -7.362222\times 10^2 \; , 
$$
$$
a_5 = 7.572929\times 10^3 \; , \qquad a_6 = - 8.273669\times 10^4 \; , \qquad 
a_7 = 9.428034\times 10^5 \; , 
$$
$$
a_8 = -1.108358\times 10^7 \; , \qquad a_9 = 1.334636\times 10^8 \; , \qquad 
a_{10} = -1.637996 \times 10^9 \; .
$$
The large $z$ limit is
\be
\label{51}
 f(z) \simeq 1.1284 z^{1/4}\qquad ( z \ra \infty) \; .
\ee
Defining the factor approximants, we find the limiting behaviour
\be
\label{52}
f_k^*(z) \simeq B_k z^{\gm_k} \qquad ( z  \ra \infty ) \;  .
\ee
The results for the amplitude and exponent are shown in Table 4. Again numerical 
convergence is clearly seen.  

\begin{table}
\caption{Amplitudes and exponents, with the related errors, for the 
large-variable limit of the function $f(z)$ for the finite-lattice Scwinger model, 
predicted by self-similar factor approximants.}
\vskip 3mm
\label{Table 4}
\renewcommand{\arraystretch}{1.25}
\centering
\begin{tabular}{|c|c|c|c|c|}
\hline
$k$ & $B_k$  &  $\ep(B_k)\%$  &  $\gm_k$  &  $\ep(\gm_k)\%$ \\ \hline
2   & 1.513  &  34.1      &  0.167 &  $-$33.0    \\
4   & 1.532  &  35.8      &  0.185 &  $-$26.2     \\
6   & 1.530  &  35.6      &  0.193 &  $-$22.7     \\
8   & 1.523  &  35.0      &  0.198 &  $-$20.7     \\
10  & 1.519  &  34.6      &  0.200 &  $-$19.8    \\   \hline
\end{tabular}
\end{table}

\section{Extrapolation with small number of terms}

It is important that self-similar factor approximants can extrapolate the series, 
derived for asymptotically small variables, to the large-variable limit, achieving 
good accuracy already for low-order approximants.

\subsection{Ground-state energy of Schwinger model in continuum limit}

The continuum limit for the ground-state energy of the Schwinger model, corresponding
to a vector boson, can be found \cite{Carrol_32,Vary_33,Adam_34,Striganesh_35} as an 
expansion in powers of the dimensionless variable $x = m/g$, where $m$ is the electron
mass and $g$ is the coupling parameter,
\be
\label{53} 
\frac{E(x)}{g} \simeq 0.5642 - 0.219 x + 0.1907 x^2 \qquad ( x \ra 0 ) \; .
\ee
A small $x \ra 0$ implies strong coupling $g \ra \infty$. In the opposite limit of 
weak coupling, hence large $x$, one has \cite{Striganesh_35,Hamer_36,Coleman_37,Hamer_38}
\be
\label{54}
 \frac{E(x)}{g} \simeq 0.6417 x^{-1/3}\qquad ( x \ra \infty) \;  .
\ee
The number of small $x$ terms is sufficient for constructing the sole factor 
approximant 
\be
\label{55}
\frac{E_2^*(x)}{g}  = \frac{0.5642}{(1+1.35339 x)^{0.286805}} \;  .
\ee
The large-$x$ limit of the latter becomes
\be
\label{56}
\frac{E_2^*(x)}{g} \simeq 0.5173 x^{-0.287}\; .
\ee
The related percentage errors are
$$
\ep(B_2) = - 17.8\% \; , \qquad \ep(\gm_2) = -13.9 \%   .
$$

\subsection{Ground-state energy of harmonium atom}

There exist finite quantum systems of different nature, but allowing for similar 
mathematical modeling \cite{Birman_39}. One of such models is the model of harmonium 
atom described by the Hamiltonian
\be
\label{57}
H = \sum_{i=1}^N \left( -\; \frac{\nabla_i^2}{2} + 
\frac{\om^2}{2}\; r_i^2 \right) + 
\frac{1}{2} \sum_{i\neq j}^N \frac{1}{|\br_i-\br_j|} \;  .
\ee
This model provides a reasonable description for several finite quantum 
systems, such as quantum dots, trapped ions, atomic nuclei, and metallic grains 
\cite{Birman_39}, because of which it has been intensively studied 
\cite{Cioslowski_40,Cioslowski_41,Cioslowski_42,Cioslowski_43,Cioslowski_44}. 

At a shallow harmonic potential, the ground-state energy of a two-electron harmonium 
atom can be presented \cite{Cioslowski_40} as an expansion in powers of the frequency 
$\omega$, 
\be
\label{58}
E(\om) \simeq \frac{3}{2^{4/3}}\; \om^{2/3} + \frac{3+\sqrt{3}}{2}\; \om
+ \frac{7}{36\cdot2^{2/3}}\; \om^{4/3} \qquad ( \om \ra 0 ) \;  .
\ee
For a rigid potential, one has
\be
\label{59}
E(\om) \simeq 3\om \qquad (\om \ra \infty ) \; .
\ee

Again, there are enough terms only for the construction of the second-order factor 
approximant
\be
\label{60}
E_2^*(\om) = \frac{3}{2^{4/3}} \; \om^{2/3} \; 
\left( 1 + 1.88379\om^{1/3} \right)^{1.05496} \; .
\ee
The predicted large-$\omega$ behaviour is
\be
\label{61}
E_2^*(\om) \simeq 2.322 \om^{1.018} \qquad ( \om \ra \infty ) \; ,
\ee
which gives the errors
$$
\ep(B_2) = -22.6\% \; , \qquad \ep(\gm_2) = 1.8 \%  \; .
$$

\subsection{Anomalous dimension in supersymmetric theory}

In the $N = 4$ supersymmetric Yang-Mills theory, the planar casp anomalous dimension 
of a lightlike Wilson loop $\Gm(g)$ is a function of the coupling parameter $g$, 
whose behaviour is known for asymptotically small $g$ and in the strong-coupling 
limit \cite{Korchemsky_45,Gubser_46,Frolov_47,Kotikov_48,Beisert_49,Correa_50}. The 
weak-coupling expansion is
\be
\label{62}
\Gm(g) \simeq 4g^2 \; - \; \frac{4\pi^2}{3}\; g^4 \; + \; 
\frac{44\pi^4}{45}\; g^6   \qquad  ( g \ra 0 ) \;   ,
\ee
while the strong-coupling limit gives
\be
\label{63}
 \Gm(g) \simeq 2g \qquad ( g \ra \infty ) \; .
\ee

The factor approximant 
\be
\label{64}
\Gm_2^*(g) = \frac{4g^2}{(1+ 11.1856 g^2)^{0.294118}}
\ee
predicts the strong-coupling limit 
\be
\label{65}
 \Gm_2^*(g) \simeq 1.966 g^{1.412} \qquad ( g \ra \infty ) \;  .
\ee 
This corresponds to the percentage errors
$$
\ep(B_2) = -1.7\% \; , \qquad \ep(\gm_2) = 41.2 \%  \;   .
$$

\subsection{Expansion factor of polymer chain}

Polymers have an extremely wide region of applicability, because of which their 
study has always been a rather hot topic \cite{Isihara_51}. The expansion factor 
of a polymer chain, as a function of the coupling parameter 
\cite{Muthukumar_52,Muthukumar_53} has the weak-coupling expansion 
\be
\label{66}
 \al(g) \simeq 1 + \sum_{n=1}^6 a_n g^n \qquad ( g \ra 0 ) \;  ,
\ee
with the fast growing coefficients
$$
a_1 = \frac{4}{3} \; , \qquad a_2 = - 2.075385396 \; , \qquad 
a_3 = 6.296879676 \; ,
$$
$$
a_4 = -25.05725072 \; , \qquad a_5 = 116.134785 \; , \qquad 
a_6 = -594.71663 \; .
$$
The strong-coupling limit \cite{Muthukumar_52,Muthukumar_53} is
\be
\label{67}
  \al(g) \simeq 1.531 g^{0.3544}  \qquad  ( g \ra \infty ) \; .
\ee

Composing the factor approximants, we find the predicted strong-coupling limit
\be
\label{68}
\al_k^*(g) \simeq B_k g^{\gm_k} \qquad ( g \ra \infty ) \;  ,
\ee
with the amplitudes and exponents presented in Table 5, together with the related 
percentage errors. Again we observe good numerical convergence.

\begin{table}
\caption{Amplitudes and exponents, with the related errors, for the 
strong-coupling limit of the polymer chain expansion factor, predicted by 
self-similar factor approximants.}
\vskip 3mm
\label{Table 5}
\renewcommand{\arraystretch}{1.25}
\centering
\begin{tabular}{|c|c|c|c|c|}
\hline
$k$ & $B_k$  &  $\ep(B_k)\%$  &  $\gm_k$  &  $\ep(\gm_k)\%$ \\ \hline
2   & 1.564  &  2.2      &  0.300 &  $-$15.4    \\
4   & 1.560  &  1.9      &  0.340 &  $-$4.1     \\
6   & 1.551  &  1.3      &  0.348 &  $-$1.9     \\  \hline
\end{tabular}
\end{table}

\section{Conclusion}

We have presented a method that makes it possible to extrapolate small-variable 
asymptotic expansions to the region of asymptotically large variables. This region is 
interesting because of two reasons. First, mathematically it is the most difficult for 
extrapolations that use only the coefficients of small-variable expansions. Second, 
the behaviour of the sought quantities in this region is often of high physical interest. 

The described method is based on self-similar approximation theory that combines the ideas
of dynamical theory, optimal control theory, and renormalization-group theory. It is the 
use of these techniques that makes the self-similar approximation theory a powerful tool
for extrapolating asymptotic series over small variables to the region of large variables.      
 
By several examples, where a number of expansion terms are available, we have 
demonstrated the numerical convergence for the sequences of approximations for 
large-variable amplitudes and exponents. The method is shown to give reasonable 
approximations even when extrapolating small-variable expansions with just a few terms.


\begin{thebibliography}{99}

\bibitem{Baker_1}
G.A. Baker and P. Graves-Moris, 
{\it Pad\'{e} Approximants} (Cambridge University Press, Cambridge, 1996).

\bibitem{Baker_2}
G.A. Baker and J.L. Gammel, 
{\it J. Math. Anal. Appl.} {\bf 2}, 21 (1961).

\bibitem{Gluzman_3}
S. Gluzman and V.I. Yukalov,  
{\it Eur. J. Appl. Math.} {\bf 25}, 595 (2014).

\bibitem{Gluzman_4}
S. Gluzman and V.I. Yukalov,  
{\it Eur. Phys. J. Plus} {\bf 131}, 340 (2016).

\bibitem{Hardy_5}
G.H. Hardy,
{\it Divergent Series} (AMS, Chelsea, Rhode Island, 1992).

\bibitem{Weinberg_6}
S. Weinberg, 
{\it The Quantum Theory of Fields} (Cambridge University, Cambridge, 2005).

\bibitem{He_7}
J.H. He,
{\it Int. J. Mod. Phys. B} {\bf 20}, 1141 (2006).

\bibitem{Yukalov_8}
V.I. Yukalov,
{\it Int. J. Mod. Phys. B} {\bf 3}, 1691 (1989).

\bibitem{Yukalov_9}
V.I. Yukalov,
{\it Phys. Rev. A} {\bf 42}, 3324 (1990).

\bibitem{Yukalov_10}
V.I. Yukalov, 
{\it J. Math. Phys.} {\bf 32}, 1235 (1991).

\bibitem{Yukalov_11}
V.I. Yukalov, 
{\it J. Math. Phys.} {\bf 33}, 3994 (1992).

\bibitem{Yukalov_12}
V.I. Yukalov,
{\it Int. J. Mod. Phys. B} {\bf 7}, 1711 (1993).

\bibitem{Yukalov_13}
V.I. Yukalov and E.P. Yukalova,
{\it Int. J. Mod. Phys. B} {\bf 7}, 2367 (1993).

\bibitem{Halmos_14}
P.R. Halmos,
{\it Lectures on Ergodic Theory} (Japan Math. Soc., Tokyo, 1956).

\bibitem{Ladyzhenskaya_15}
O. Ladyzhenskaya,
{\it Attractors for Semigroups and Evolution Equations} 
(Cambridge University, Cambridge, 1991).

\bibitem{Lichtenberg_16}
A.J. Lichtenberg and M.A. Liberman,
{\it Regular and Chaotic Dynamics} (Springer, New York, 1992).

\bibitem{Yukalov_17}
V.I. Yukalov,
{\it Phys. Part. Nucl.} {\bf 50}, 141 (2019).

\bibitem{Foulds_18}
L.R. Foulds,
{\it Optimization Techniques} (Springer, New York, 1981). 

\bibitem{Hocking_19}
L.M. Hocking,
{\it Optimal Control} (Clarendon, Oxford, 1991).

\bibitem{Adhikari_54}
S.K. Adhikari, T. Frederico and I.D. Goldman,
{\it Phys. Rev. Lett.} {\bf 74}, 487 (1995).

\bibitem{Adhikari_55}
S.K. Adhikari and A. Ghosh,
{\it J. Phys. A} {\bf 30}, 6553 (1997). 

\bibitem{Yukalov_20}
V.I. Yukalov, S. Gluzman and D. Sornette, 
{\it Physica A} {\bf 328}, 409 (2003).

\bibitem{Gluzman_21}
S. Gluzman, V.I. Yukalov and D. Sornette, 
{\it Phys. Rev. E} {\bf 67}, 026109 (2003).

\bibitem{Yukalov_22}
V.I. Yukalov and E.P. Yukalova, 
{\it Eur. Phys. J. B} {\bf 55}, 93 (2007).

\bibitem{Gluzman_23}
S. Gluzman and V.I. Yukalov,
{\it Int. J. Mod. Phys. B} {\bf 33}, 1950353 (2019). 

\bibitem{Lang_24}
S. Lang, 
{\it Algebra} (Addison-Wesley, Reading, 1984).

\bibitem{Barnsley_25}
M.F. Barnsley,
{\it Fractal Transform} (AK Peters Ltd., Natick, 1994).

\bibitem{Yukalov_26}
V.I. Yukalov and E.P. Yukalova,
{\it Chaos Solit. Fract.} {\bf 14}, 839 (2002).

\bibitem{Yukalov_27}
V.I. Yukalov and E.P. Yukalova,
{\it Laser Phys. Lett.} {\bf 14}, 073001 (2017).

\bibitem{Bender_28}
C.M. Bender and T.T. Wu, 
{\it Phys. Rev.} {\bf 184}, 1231 (1969).

\bibitem{Hioe_29}
F.T. Hioe, D. MacMillen and E.W. Montroll,
{\it Phys. Rep.} {\bf 43}, 305 (1978).

\bibitem{Schwinger_30}
J. Schwinger, 
{\it Phys. Rev.} {\bf 128}, 2425 (1962).

\bibitem{Banks_31}
T. Banks, L. Susskind and J. Kogut, 
{\it Phys. Rev. D} {\bf 13},  1043 (1976).

\bibitem{Carrol_32}
A. Carrol, J. Kogut, D.K. Sinclair and L. Susskind, 
{\it Phys. Rev. D} {\bf 13}, 2270 (1976).

\bibitem{Vary_33} 
J.P. Vary, T.J. Fields and H.J. Pirner, 
{\it Phys. Rev. D} {\bf 53}, 7231 (1996).

\bibitem{Adam_34}
C. Adam, 
{\it Phys. Lett. B} {\bf 382},  383 (1996).

\bibitem{Striganesh_35}
P. Striganesh, C.J. Hamer and R.J. Bursill, 
{\it Phys. Rev. D} {\bf 62}, 034508 (2000).

\bibitem{Hamer_36}
C.J. Hamer, Z. Weihong and J. Oitmaa, 
{\it Phys. Rev. D} {\bf 56}, 55 (1997).

\bibitem{Coleman_37}
S. Coleman, 
{\it Ann. Phys. (N.Y.)} {\bf 101}, 239 (1976).

\bibitem{Hamer_38}
C.J. Hamer,  
{\it Nucl. Phys. B} {\bf 121}, 159 (1977).

\bibitem{Birman_39}
J.L. Birman, R.G. Nazmitdinov and V.I. Yukalov, 
{\it Phys. Rep.} {\bf 526}, 1 (2013).

\bibitem{Cioslowski_40}
J. Cioslowski, 
{\it J. Chem. Phys.} {\bf 136}, 044109 (2012).

\bibitem{Cioslowski_41}
J. Cioslowski, K. Strasburger and E. Matito, 
{\it J. Chem. Phys.} {\bf 136}, 194112 (2012).

\bibitem{Cioslowski_42}
J. Cioslowski and J. Albin, 
{\it J. Chem. Phys.} {\bf 139}, 114109 (2013).

\bibitem{Cioslowski_43}
J. Cioslowski, 
{\it J. Chem. Phys.} {\bf 139}, 220148 (2013).

\bibitem{Cioslowski_44}
J. Cioslowski, K. Strasburger and E. Matito, 
{\it J. Chem. Phys.} {\bf 141}, 044128 (2014).

\bibitem{Korchemsky_45}
G.P. Korchemsky and A.V. Radyushkin, 
{\it Nucl. Phys. B} {\bf 283}, 342 (1987).

\bibitem{Gubser_46}
S.S. Gubser, I.R. Klebanov and A.M. Polyakov, 
{\it Nucl. Phys. B} {\bf 636}, 99 (2002).

\bibitem{Frolov_47}
S. Frolov and A.A. Tseytlin, 
{\it J. High Energy Phys.} {\bf 06}, 007 (2002).

\bibitem{Kotikov_48}
A.V. Kotikov, L.N. Lipatov, A.I. Onishchenko and V.N. Velizhanin, 
{\it Phys. Lett. B} {\bf 595}, 521 (2004).

\bibitem{Beisert_49}
N. Beisert, B. Eden and M. Staudacher, 
{\it J. Stat. Mech.} {\bf 2007}, 01021 (2007).

\bibitem{Correa_50}
D. Correa, J. Henn, J. Maldacena and A. Sever, 
{\it J. High Energy Phys.} {\bf 05}, 098 (2012).

\bibitem{Isihara_51}
A. Isihara,
{\it Prog. Theor. Phys.} {\bf 97}, 709 (1997).

\bibitem{Muthukumar_52}
M. Muthukumar and B.G. Nickel,
{\it J. Chem. Phys.} {\bf 80}, 5839 (1984).

\bibitem{Muthukumar_53}
M. Muthukumar and B.G. Nickel,
{\it J. Chem. Phys.} {\bf 86}, 460 (1987).


\end{thebibliography}
\end{document}